\documentclass[notitlepage,twocolumn,pra]{revtex4} 

\newcounter{mycounter}

\newcounter{reference}
\newenvironment{Reference}
   {\begin{list}{\bf \arabic{reference}.}{\usecounter{reference}%
       \setlength{\leftmargin}{2.5em}%
       \setlength{\itemsep}{1pt}%
       \setlength{\parsep}{0pt}%
       \setlength{\parskip}{0pt}%
       \setlength{\parindent}{0pt}%
       \setlength{\itemindent}{0pt}%
      }%
   }%
   {\end{list}}

\newcommand{\referencefont}{\scriptsize} 

\newcommand{\bref}{\referencefont\begin{Reference}\setcounter{reference}{\value{mycounter}}}
\newcommand{\eref}{\setcounter{mycounter}{\value{reference}}\end{Reference}\normalsize}
\begin{document}

\title{Resource Letter BEC-1: Bose-Einstein Condensates in Trapped Dilute Gases}

\author{David S. Hall}
\affiliation{Department of Physics, Amherst College, Amherst,
Massachusetts 01002--5000 USA}%

\date{November 29, 2002}

\begin{abstract}

This Resource Letter provides a guide to the literature on
Bose-Einstein condensation in trapped dilute gases. Journal
articles and books are cited for the following topics: history,
technological advances, condensates as quantum fluids, effects of
interatomic interactions, condensates as matter waves, condensate
optics, multiple condensates, lower dimensions, spectroscopy and
precision measurement, entanglement, and cosmology.

\end{abstract}

\maketitle


\section{INTRODUCTION}

It is amusing to note that the field of Bose-Einstein condensation
in dilute gases, extremely rich and vibrant today, traces its
origins to a rejected paper in 1924~[\ref{subtle}]. The author of
the paper, S.~N. Bose, resubmitted his paper on what would become
known as ``Bose statistics'' to no less an authority than Albert
Einstein, who arranged for its publication. In a subsequent
trilogy of papers, published in 1925, Einstein considered the
implications of Bose statistics for an ideal gas and determined
that at sufficiently low temperatures and high densities the gas
atoms would collect predominantly in the ground state of the
system. This macroscopic occupation of a single-particle state,
now known as Bose-Einstein condensation (BEC), is peculiar to
systems of particles of integer spin (bosons), being forbidden in
systems of half-integer spin (fermions) by the Pauli exclusion
principle.

The condition for the appearance of a Bose-Einstein condensate may
be stated in terms of phase-space density $\mathcal{D}$:
\begin{equation}
\mathcal{D} = n \Lambda_\mathrm{dB}^3,
\end{equation}
where $n$ is the particle density. Here
\begin{equation}
\Lambda_\mathrm{dB}=\frac{h}{\sqrt{2\pi m k_B T}}
\end{equation}
is the thermal de Broglie wavelength, a measure of the average
size of the atomic wavepacket for an atom of mass $m$ in a sample
at temperature $T$. The quantity $\Lambda_\mathrm{dB}^3$ is
related to the ``volume'' such an atom occupies, and its product
with the particle density is a dimensionless number representing
the degree of overlap of the individual atomic wavepackets. When
$\mathcal{D}$ becomes greater than a number of order~1, the
wavepackets begin to overlap and the effects of quantum degeneracy
begin to emerge.

Seventy years would pass before Bose-Einstein condensates were
experimentally realized in weakly-interacting, near-ideal
gases~[\ref{jila1},\ref{rice1},\ref{mit1}]. In the meantime,
macroscopic quantum behavior such as superfluidity in liquid
helium and superconductivity in metals were convincingly related
to the existence of Bose-Einstein condensates (see
Ref.~[\ref{bec}] and references therein for a review). Theoretical
progress had been steady but difficult owing to the complicated
nature of the interactions between the particles of which these
condensates are composed.

The experimental search for BEC in a weakly-interacting dilute gas
began with spin-polarized hydrogen~[\ref{silvera},\ref{greytak}],
which was predicted to remain a gas all the way to zero
temperature~[\ref{hecht},\ref{stwalley}]. Two techniques that
played central roles in this and subsequent work with the dilute
alkalis (see Ref.~[\ref{tna}] for a review) are the magnetic trap,
which confines atoms by their magnetic moments in strongly
inhomogenous magnetic fields, and evaporative
cooling~[\ref{evap}], whereby the gas sample is cooled by
preferentially permitting the most energetic atoms to escape.

The invention of the magneto-optical trap (MOT) in 1987 heralded
the beginning of the modern era of BEC research in the
alkalis~[\ref{tna}]. A MOT permitted production of a cold atomic
sample (10-100~$\mu$K) at reasonable densities (up to
$10^{12}~\mathrm{cm}^{-3}$) consisting of atoms with convenient
atomic transitions, such as the alkalis and metastable excited
noble gases. Progress in further cooling and increasing the
density of the sample was hampered, however, by the rescattering
of the very light used to cool the sample.

The next step was to turn off the light and proceed in the dark:
MOT-cooled atoms were loaded into a trap and confined by the
interaction between their magnetic moments and an inhomogenous
applied magnetic field. Once inside this potential, the atoms
could be further cooled using evaporative cooling~[\ref{evap}].
This process relied on driving transitions between trapped and
untrapped internal states of the atom in a spatially-selective
fashion --- that is, selecting for removal those with the largest
amplitude orbits, and hence the largest energies. The remaining
atoms rethermalize through collisions at decreasing temperatures
until they reach the phase transition and begin to pile up in the
ground state of the trap.

The emergence of the condensate was first observed at JILA in
1995~[\ref{jila1}], followed closely by reports at Rice
University~[\ref{rice1}] and MIT~[\ref{mit1}]. Experimental and
theoretical work proceeded swiftly, aided by two considerations.
On the experimental side, the data collection and analysis
generally involved imaging the condensates with a camera,
permitting nearly direct visualization of the macroscopically
occupied quantum state. This led to some stunning images, such as
those of the interference fringes that appear between two
overlapping condensates~[\ref{andrews}]. At the same time, the
theoretical analysis remained fairly simple, with qualitatively
new features arising from the inhomogeneity of the trapping
potential.

The Nobel Prize in 2001 was awarded to Eric Cornell, Wolfgang
Ketterle, and Carl Wieman for their work on
BEC~[\ref{nobel1},\ref{nobel2},\ref{nobel}]. More than fifty
groups worldwide now are producing dilute-gas condensates in a
variety of different atoms, using an ever-expanding collection of
tricks and techniques~[\ref{becbib}]. An enormous amount of
theoretical work has accompanied and driven the experimental
progress, with tendrils expanding into many branches of physics.

One consequence of these rapid developments is that the current
literature on BEC in dilute gases can be overwhelming to novices
and experts alike. This Resource Letter aims to complement the
several theoretical review articles that already have appeared by
maintaining a slight bias towards experimental work. Even so, the
references are necessarily incomplete. The fantastic pace of
progress in this field, moreover, insures that this certainly will
not be the last Resource Letter on BEC in the weakly interacting
dilute Bose gases.

\section{JOURNALS}

The majority of the technical articles on BEC in dilute gases are
found in the first four journals below. Technical articles in both
\textit{Nature} and \textit{Science} often are accompanied by more
general descriptions of the work and its context for a wider
audience.

\vspace*{\baselineskip}
\setlength{\parindent}{0pt}

{\it Physical Review Letters}\\
{\it Physical Review A}\\
{\it Nature}\\
{\it Science}\\
{\it Europhysics Letters}\\
{\it Journal of Physics B}\\
{\it Journal of Low Temperature Physics}\\
{\it Physics Today}\\
{\it Physics World}\\
{\it Scientific American}\\
{\it Optics Express} (electronically published at
\texttt{www.opticsexpress.org})

\setlength{\parindent}{1em}

\section{BOOKS}

\referencefont

\begin{Reference}

\item \label{bec}
{\bf Bose-Einstein Condensation,} edited by A. Griffin, D. W.
Snoke, and S. Stringari, (Cambridge University Press, Cambridge,
1995). A general guide to the various manifestations of
Bose-Einstein condensation, with some sections on BEC in dilute
gases. (I)

\item \label{varenna} {\bf Proceedings, Enrico Fermi International Summer School on
Bose-Einstein Condensation in Atomic Gases,} Varenna, Italy,
edited by M.~Inguscio, S.~Stringari, and C.~E. Wieman (IOS Press,
Washington, 1999). These proceedings from the Varenna Summer
School on Bose-Einstein condensation contain the most important
collection of review articles to date. (I)

\item
{\bf Bose-Einstein Condensates and Atom Lasers,} S.~Martellucci,
edited by A.~N. Chester, A.~Aspect, and M.~Inguscio (Kluwer
Academic/Plenum Publishers, New York, 2000).  These are the
proceedings from the Erice lectures on Bose-Einstein condensation.
(I)

\item
{\bf Introduction to Statistical Physics,} K.~Huang (Taylor and
Francis, New York, 2001). Chapter~11 of this textbook contains a
concise introduction to Bose-Einstein condensation. (I)

\item {\bf Bose-Einstein Condensation in Dilute Gases,} C. J.
Pethick and H. Smith (Cambridge University Press, Cambridge,
2002). This textbook provides a comprehensive overview to the
field. (I)

\eref

\section{REVIEW ARTICLES}

There already are many review articles on Bose-Einstein
condensation for both the novice and expert. Each article listed
below covers several different aspects of BEC. Additional articles
that provide comprehensive reviews on single topics are listed at
the beginning of the relevant sections of this Resource Letter.

\bref

\item
``Bose-Einstein Condensation with Evaporatively Cooled Atoms,'' K.
Burnett, Contemp. Phys. {\bf 37}, 1--14 (1996). (I)

\item
``The {R}ichtmyer Memorial Lecture: {B}ose-{E}instein Condensation
in an Ultracold Gas,'' C.~E. Wieman, Am. J. Phys. {\bf 64},
847--855 (1996). (E)

\item
\textit{Special Issue on Bose-Einstein Condensation,} edited by K.
Burnett, M.~Edwards, and C.~W. Clark, J. Res. NIST {\bf 101},
419--600 (1996). (I)

\item
``{B}ose-{E}instein Condensation,'' C.~G. Townsend, W.~Ketterle,
and S.~Stringari, Phys. World {\bf 10}, 29--34 (1997). (E)

\item
``Bose-Einstein Condensation,'' I.~F. Silvera, Am. J. Phys. {\bf
65}, 570--574 (1997). This review contains some simple
textbook-style problems. (E)

\item
``The {B}ose-{E}instein Condensate,'' E.~A. Cornell and C.~E.
Wieman, Sci. Am. {\bf 278}(3), 40--45 (March 1998). (E)

\item
``The Physics of Trapped Dilute-Gas Bose-Einstein Condensates,''
A.~S. Parkins and D.~F. Walls, Phys. Rep. {\bf 303}, 1--50 (1998).
(I)

\item
``Experimental Studies of {B}ose-{E}instein Condensation,''
W.~Ketterle, Phys. Today {\bf 52}(12), 30--35 (December 1999). (E)

\item
``The Theory of {B}ose-{E}instein Condensation of Dilute Gases,''
K.~Burnett, M.~Edwards, and C.~W. Clark, Phys. Today {\bf 52}(12),
37--42 (December 1999). (E)

\item
``Bose Condensates Make Quantum Leaps and Bounds,'' Y. Castin, R.
Dum, and A. Sinatra, Phys. World {\bf 12}(8), 37--42 (August
1999). (E)

\item ``Theory of {B}ose-{E}instein Condensation in Trapped
Gases,'' F. Dalfovo, S. Giorgini, L.~P. Pitaevskii, and S.
Stringari, Rev. Mod. Phys. {\bf 71}, 463--512 (1999). (I)

\item
``{B}ose-{E}instein Condensation in the Alkali Gases: Some
Fundamental Concepts,'' A.~J. Leggett, Rev. Mod. Phys. {\bf 73},
307--356 (2001). (I)

\item ``Bose-Einstein Condensation of Trapped Atomic Gases,'' Ph.
W. Courteille, V. S. Bagnato, and V. I. Yukalov, Laser Phys. {\bf
11}, 659--800 (2001). (A)

\item
``Bose-Einstein Condensation of Atomic Gases,'' J.~R. Anglin and
W.~Ketterle, Nature {\bf 416}, 211--218 (2002). This is one of six
articles on the physics of cold atoms in this issue of Nature. (E)

\item \label{nobel1}
``Nobel Lecture: {B}ose-{E}instein Condensation in a Dilute Gas,
the First 70 Years and Some Recent Experiments,'' E.~A. Cornell
and C.~E. Wieman, Rev. Mod. Phys. {\bf 74}, 875--893 (2002). (E)

\item \label{nobel2}
``Nobel Lecture: When Atoms Behave as Waves: Bose-Einstein
Condensation and the Atom Laser,'' W.~Ketterle, Rev. Mod. Phys.
{\bf 74}, 1131--1151 (2002). (E)

\eref

\section{WEB SITES}

\bref

\item \label{becbib}
The BEC Online Bibliography at Georgia Southern University: \\
\texttt{http://amo.phy.gasou.edu:80/bec.html/bibliography.html} \\
contains links to a large number of preprints and up-to-date
information on the current status of the field.

\item \label{gasou}
The Electronic Preprint Archive (\texttt{http://cul.arxiv.org/})
also contains preprints. These papers typically are found in the
\texttt{cond-mat} section.

\item
The Physics 2000 web site,
\texttt{http://www.colorado.edu/physics/2000/bec/index.html}, has
descriptions and information for the general audience, and some
interesting computer simulations.

\item \label{nobel}
Eric~A. Cornell, Wolfgang Ketterle, and Carl~E. Wieman won the
2001 Nobel Prize in physics for ``for the achievement of
Bose-Einstein condensation in dilute gases of alkali atoms, and
for early fundamental studies of the properties of the
condensates.'' More details are available at the official Nobel
web site
\texttt{http://www.nobel.se/physics/laureates/2001/index.html}.

\item
Recent experiments from the groups of E.~A. Cornell and C.~E.
Wieman are detailed on the JILA web site
\texttt{http://jilawww.colorado.edu/bec/}. The JILA group also
maintains at this web site a comprehensive bibliography of
published works concerning BEC, complementary to the preprint
listing in Ref.~[\ref{gasou}].

\item
Recent experiments from the group of W.~Ketterle are discussed on
the MIT web site \texttt{http://cua.mit.edu/ketterle\_group/}.

\item
There has been considerable experimental and theoretical work on
BEC at NIST, as documented on their web site
\texttt{http://bec.nist.gov/}.

\eref

\section{PARTICULAR TOPICS}

\subsection{History}

\subsubsection{First Inklings and Attempts}

Additional historical perspectives can be found in articles in
Refs.~[\ref{bec},\ref{varenna}] as well as Ref.[\ref{nobel}].

\bref

\item \label{subtle}
{\bf Subtle is the Lord ... The Science and the Life of Albert
Einstein,} A.~Pais (Oxford University Press, New York, 1982).
Chapters~19 and~23 are of particular relevance. (E)

\item
``Quantentheorie des Einatomigen Idealen Gases. Zweite
Abhandlung,'' A. Einstein, Sitzungsberichte der Preussischen
Akademie der Wissenschaften I, 3--14 (1925). This is the paper in
which Einstein establishes the existence of a condensate in an
ideal gas. Although it is written in German, it should eventually
appear in translation as a part of Einstein's collected works. (E)

\item
``The $\lambda$-Phenomenon of Liquid Helium and the Bose-Einstein
Degeneracy,'' F. London, Nature {\bf 141}, 643--644 (1938). This
foundation paper is available online through the Nature Publishing
Group Physics Portal (\texttt{http://www.nature.com/physics/}).
(E)

\item \label{hecht}
``The Possible Superfluid Behaviour of Hydrogen Atom Gases and
Liquids,'' C.~E. Hecht, Physica {\bf 25} 1159--1161 (1959). (I)

\item \label{stwalley}
``Possible `New' Quantum Systems,'' W.~C. Stwalley and L.~H.
Nosanow, Phys. Rev. Lett. {\bf 36}, 910--913 (1976). (I)



\item \label{silvera}
``The Stabilization of Atomic Hydrogen,'' I.~F. Silvera and
J.~Walraven, Sci. Am. {\bf 246}(1), 66--74 (January 1982). (E)

\item \label{greytak}
``Lectures on Spin-Polarized Hydrogen,'' T.~J. Greytak and
D.~Kleppner, in {\bf New Trends in Atomic Physics, Vol.~2,} edited
by G.~Grynberg and R.~Stora (North-Holland, New York, 1984),
pp.~1125--1230. This set of lecture notes is from the 38th
Les~Houches summer school in 1982. (I)


\eref

\subsubsection{First Realizations}

Dilute-gas Bose-Einstein condensates have been realized in
${}^{87}$Rb, ${}^{7}$Li, ${}^{23}$Na, ${}^{1}$H, ${}^{85}$Rb,
${}^{41}$K, metastable ${}^{4}\mathrm{He}^*$, and ${}^{133}$Cs.
The following papers give accounts of the production of the first
condensates in these systems --- an introduction to the BEC
``family tree.''

\bref

\item \label{jila1}
``Observation of {B}ose-{E}instein Condensation in a Dilute Atomic
Vapor,'' M.~H. Anderson, J.~R. Ensher, M.~R. Matthews, C.~E.
Wieman, and E.~A. Cornell, Science {\bf 269}, 198--201 (1995). (I)

\item \label{rice1}
``Evidence of {B}ose-{E}instein Condensation in an Atomic Gas with
Attractive Interactions,'' C.~C. Bradley, C.~A. Sackett, J.~J.
Tollett, and R.~G. Hulet, Phys. Rev. Lett. {\bf 75}, 1687--1690
(1995); {\it ibid.} {\bf 79}, 1170 (1997). (I)

\item \label{mit1}
``{B}ose-{E}instein Condensation in a Gas of Sodium Atoms,'' K.~B.
Davis, M.-O. Mewes, M.~R. Andrews, N.~J. van Druten, D.~S. Durfee,
D.~M. Kurn, and W.~Ketterle, Phys. Rev. Lett. {\bf 75}, 3969--3973
(1995). (I)

\item
``Bose-{E}instein Condensation of Atomic Hydrogen,'' D.~G. Fried,
T.~C. Killian, L.~Willmann, D.~Landhuis, S.~C. Moss, D.~Kleppner,
and T.~J. Greytak, Phys. Rev. Lett. {\bf 81}, 3811--3814 (1998).
(I)

\item \label{rubidium}
``Stable $^{85}${R}b {B}ose-{E}instein Condensates with Widely
Tunable Interactions,'' S.~L. Cornish, N.~R. Claussen, J.~L.
Roberts, E.~A. Cornell, and C.~E. Wieman, Phys. Rev. Lett. {\bf
85}, 1795--1798 (2000). (I)

\item
\label{Modugno2001} ``Bose-{E}instein Condensation of Potassium
Atoms by Sympathetic Cooling,'' G.~Modugno, G.~Ferrari, G.~Roati,
R.~J. Brecha, A.~Simoni, and M.~Inguscio, Science {\bf 294},
1320--1322 (2001). (I)

\item
``A {B}ose-{E}instein Condensate of Metastable Atoms,'' A.~Robert,
O.~Sirjean, A.~Browaeys, J.~Poupard, S.~Nowak, D.~Boiron, C.~I.
Westbrook, and A.~Aspect, Science {\bf 292}, 461--464 (2001). (I)

\item
``{B}ose-{E}instein Condensation of Metastable Helium,'' F.~P.
Dos~Santos, J.~L{\'e}onard, J.~Wang, C.~J. Barrelet, F.~Perales,
E.~Rasel, C.~S. Unnikrishnan, M.~Leduc, and C.~Cohen-Tannoudji,
Phys. Rev. Lett. {\bf 86}, 3459--3463 (2001). (I)

\item \label{cesium}
``{B}ose-{E}instein Condensation of Cesium,''  T.~Weber,
J.~Herbig, M.~Mark, H.-C. N{\"a}gerl, and R.~Grimm, Science {\bf
299}, 232--235 (2003). (I)

\eref

\subsubsection{Prospects in Other Atoms}

Experimenters are working hard on extending the dilute-gas
techniques to other atoms, as this brief list suggests.

\bref

\item
``Prospects for Bose-Einstein Condensation of Metastable Neon
Atoms,'' H.~C.~W. Beijerinck, E.~J.~D. Vredenbregt, R.~J.~W. Stas,
M.~R. Doery, and J.~G.~C. Tempelaars, Phys. Rev. A {\bf 61},
023607/1--15 (2000). (I)

\item
``Optical-Dipole Trapping of Sr Atoms at a High Phase-Space
Density,'' T.~Ido, Y.~Isoya, and H.~Katori, Phys. Rev. A {\bf 61},
061403/1--4 (2000). (I)

\item
``Evaporative Cooling of Atomic Chromium,'' J.~D. Weinstein,
R.~deCarvalho, C.~I. Hancox, and J.~M. Doyle, Phys. Rev. A {\bf
65}, 021604/1--4 (2002). (I)

\eref

\subsection{Technological Advances}

Advances in technology have led to new methods of confining atoms,
creating and imaging condensates, and transporting and guiding
them into new experimental configurations. Using the newly
available techniques, experimenters bring an exquisite level of
control and manipulation to these systems.

\bref

\item \label{tna}
``Resource Letter TNA-1: Trapping of Neutral Atoms,'' N. R.
Newbury and C. Wieman, Am. J. Phys. {\bf 64}, 18--20 (1996). This
Resource Letter contains a large number of references on the
advances that led to Bose-Einstein condensation in dilute gases.
(E)

\item \label{evap}
``Evaporative Cooling of Trapped Atoms,'' W.~Ketterle and N.~J.
Van Druten, Adv. At. Mol. Opt. Phys. {\bf 37}, 181--236 (1996).
Review article. (I)

\item
``{B}ose-{E}instein Condensation in a Tightly Confining DC
Magnetic Trap,'' M.-O. Mewes, M.~R. Andrews, N.~J. van Druten,
D.~M. Kurn, D.~S. Durfee, and W.~Ketterle, Phys. Rev. Lett. {\bf
77}, 416--419 (1996). (I)

\item
``Direct, Nondestructive Observation of a {B}ose Condensate,''
M.~R. Andrews, M.-O. Mewes, N.~J. van Druten, D.~S. Durfee, D.~M.
Kurn, and W.~Ketterle, Science {\bf 273}, 84--87 (1996). (I)

\item
\label{opticaltrap} ``Optical Confinement of a {B}ose-{E}instein
Condensate,'' D.~M. Stamper-Kurn, M.~R. Andrews, A.~P. Chikkatur,
S.~Inouye, H.-J. Miesner, J.~Stenger, and W.~Ketterle, Phys. Rev.
Lett. {\bf 80}, 2027--2030 (1998). (I)

\item
``Bose-Einstein Condensation in a Quadrupole-Ioffe-Configuration
Trap,'' T.~Esslinger, I.~Bloch, and T.~W. H{\"a}nsch, Phys. Rev. A
{\bf 58}, R2664--R2667 (1998). (I)


\item ``All-Optical Formation of an Atomic {B}ose-{E}instein
Condensate,'' M.~D. Barrett, J.~A. Sauer, and M.~S. Chapman, Phys.
Rev. Lett. {\bf 87}, 010404/1--4 (2001). (I)

\item
``Observation of Coherent Optical Information Storage in an Atomic
Medium using Halted Light Pulses,'' C.~Liu, Z.~Dutton, C.~H.
Behroozi, and L.~V. Hau, Nature {\bf 409}, 490--493 (2001). (I)

\item
``{B}ose-{E}instein Condensation on a Microelectronic Chip,''
W.~H{\"a}nsel, P.~Hommelhoff, T.~W. H{\"a}nsch, and J.~Reichel,
Nature {\bf 413}, 498--501 (2001). (I)

\item
``{B}ose-{E}instein Condensation in a Surface Microtrap,'' H.~Ott,
J.~Fortagh, G.~Schlotterbeck, A.~Grossmann, and C.~Zimmermann,
Phys. Rev. Lett. {\bf 87}, 230401/1--4 (2001). (I)

\item
``Waveguide for Bose-Einstein Condensates,'' K.~Bongs, S.~Burger,
S.~Dettmer, D.~Hellweg, J.~Arlt, W.~Ertmer, and K.~Sengstock,
Phys. Rev. A {\bf 63}, 031602/1--4 (2001). (I)

\item
``Transport of {B}ose-{E}instein Condensates with Optical
Tweezers,'' T.~L. Gustavson, A.~P. Chikkatur, A.~E. Leanhardt,
A.~G{\"o}rlitz, A.~Gupta, D.~E. Pritchard, and W.~Ketterle, Phys.
Rev. Lett. {\bf 88}, 020401/1--4 (2002). (I)

\item
``A Continuous Source of {B}ose-{E}instein Condensed Atoms,''
A.~P. Chikkatur, Y.~Shin, A.~E. Leanhardt, D.~Kielpinski,
E.~Tsikata, T.~L. Gustavson, D.~E. Pritchard, and W.~Ketterle,
Science {\bf 296}, 2193--2195 (2002). (I)

\item
``Propagation of {B}ose-{E}instein Condensates in a Magnetic
Waveguide,'' A.~E. Leanhardt, A.~P. Chikkatur, D.~Kielpinski,
Y.~Shin, T.~L. Gustavson, W.~Ketterle, and D.~E. Pritchard, Phys.
Rev. Lett. {\bf 89}, 040401/1--4 (2002). (I)

\eref

\subsection{Condensates as Quantum Fluids}

\subsubsection{Ground State}

The ground state of a weakly interacting Bose gas is discussed at
length in many standard textbooks. Experimentally realized
Bose-degenerate gases differ from their textbook counterparts
primarily in that they are trapped, and therefore inhomogeneous.
Inhomogeneity gives rise to condensation in both momentum and
position space, and permits (small) condensates to form even with
attractive interatomic interactions.

The theoretical understanding of the ground state of a Bose
condensate usually begins with the time-independent
Gross-Pitaevskii (GP) equation, or nonlinear Schr\"odinger
equation,
\begin{equation}
\label{eq.gpe} -\frac{\hbar^2}{2m}\nabla^2\Psi + V \Psi +
g|\Psi|^2\Psi = \mu \Psi
\end{equation}
for the macroscopic condensate wave function (or order parameter)
$\Psi$, where $m$ is the atomic mass, $V$ is the external trapping
potential, $\mu$ is the chemical potential, and $g$ is the
nonlinear coupling strength. (The theoretical foundations of the
GP equation are developed Refs.~[\ref{bogolubov}--\ref{gross}]
below.) The nonlinear term in the GP equation is the so-called
``mean-field'' interaction that effectively accounts for the
short-range interactions between the atoms that make up the
condensate. Remarkably, the strength of the mean-field interaction
depends upon a single parameter: the $s$-wave scattering length
$a$:
\begin{equation}
g = \frac{4\pi \hbar^2 a}{m}.
\end{equation}
Almost every aspect of the condensate depends crucially on $a$,
and articles that touch on this relationship are listed in
Sec.~\ref{Feshbach}, below.


The scattering length vanishes for Einstein's noninteracting ideal
gas ($a=0$), turning the GP equation into a linear
Schr{\"o}dinger-like equation for the macroscopic wavefunction. In
the opposite limit, that is,
\begin{equation}
\frac{8\pi a N}{a_\mathrm{HO}} \gg 1
\end{equation}
(where $a_\mathrm{HO}=\sqrt{\hbar/(m\omega)}$ for atoms in a trap
of frequency $\omega$), the mean-field term of the GP equation
dominates. One then can neglect the kinetic-energy term (first
term on the left-hand side of Eq.~\ref{eq.gpe}) to arrive at an
algebraic equation for $\Psi$. This is the Thomas-Fermi
approximation.

The GP equation is an approximation that is valid provided the gas
is dilute, that is, $n|a|^3 \ll 1$; otherwise, corrections beyond
the mean-field approximation are required.

\bref

\item \label{bogolubov}
``On the Theory of Superfluidity,'' N.~N. Bogolubov, J. Phys. USSR
{\bf 11}, 23--32 (1947). Foundation paper. (A)

\item
``Quantum-Mechanical Many-Body Problem with Hard-Sphere
Interaction,'' K.~Huang and C.~N. Yang, Phys. Rev. {\bf 105},
767--775 (1957). Foundation paper. (I)

\item
``Vortex Lines in an Imperfect Bose Gas,'' L.~P. Pitaevskii, Sov.
Phys. JETP {\bf 13}, 451--454 (1961). Foundation paper. (I)

\item \label{gross}
``Structure of a Quantized Vortex in Boson Systems,'' E.~P. Gross,
Nuovo Cimento {\bf 20}, 454--477 (1961). Foundation paper. (A)

\item
``Numerical Solution of the Nonlinear Schr{\"o}dinger Equation for
Small Samples of Trapped Neutral Atoms,'' M.~Edwards and
K.~Burnett, Phys. Rev. A {\bf 51}, 1382--1386 (1995). (I)

\item
``Ground-State Properties of Magnetically Trapped {B}ose-condensed
Rubidium Gas,'' G.~Baym and C.~J. Pethick, Phys. Rev. Lett. {\bf
76}, 6--9 (1996). (I)

\item
``Expansion of a Bose-Einstein Condensate in a Harmonic
Potential,'' M.~Holland and J.~Cooper, Phys. Rev. A {\bf 53},
R1954--R1957 (1996). (I)

\item
``Order Parameter at the Boundary of a Trapped Bose Gas,''
F.~Dalfovo, L.~Pitaevskii, and S.~Stringari, Phys. Rev. A {\bf
54}, 4213--4217 (1996). (A)

\item
``Bose-{E}instein Condensation of Lithium: {O}bservation of
Limited Condensate Number,'' C.~C. Bradley, C.~A. Sackett, and
R.~G. Hulet, Phys. Rev. Lett. {\bf 78}, 985--989 (1997). (I)

\item
``Emergence of Interaction Effects in Bose-Einstein
Condensation,'' M.~J. Holland, D.~S. Jin, M.~L. Chiofalo, and
J.~Cooper, Phys. Rev. Lett. {\bf 78}, 3801--3805 (1997). (I)

\item
``Near-Resonant Spatial Images of Confined {B}ose-{E}instein
Condensates in a 4-{D}ee Magnetic Bottle,'' L.~V. Hau, B.~Busch,
C.~Liu, Z.~Dutton, M.~M. Burns, and J.~Golovchenko, Phys. Rev. A
{\bf 58}, R54--R57 (1998). (I)

\item
``Spatial Observation of Bose-Einstein Condensation,'' B.~P.
Anderson and M.~A. Kasevich, Phys. Rev. A {\bf 59}, R938--R941
(1999). (I)

\eref

\subsubsection{Collective Excitations and Sound}

\normalsize

The dynamics of the condensate can be studied by examining its
response to perturbations of the ground state. In particular,
oscillating solutions (modes) of the time-dependent GP equation
are sought:
\begin{equation}
i\hbar \frac{\partial \Psi}{\partial t} =
-\frac{\hbar^2}{2m}\nabla^2\Psi + V \Psi + g|\Psi|^2\Psi.
\end{equation}
The problem generally is approached by linearizing the GP equation
and calculating the excitation spectrum. One also can do this by
converting the time-dependent GP equation into a set of
hydrodynamic equations for the density, $\rho=|\Psi|^2$, and the
velocity field $\mathbf{v}=\hbar \nabla \phi/m$, where $\phi$ is
the phase of the condensate order parameter: $\Psi =
\sqrt{\rho}\mathrm{e}^{i\phi}$.

For a condensate in a noninteracting (ideal) gas, the excitation
spectrum is entirely particle-like:
\begin{equation}
\label{eq.particlelike} E = \frac{\hbar^2 q^2}{2m}
\end{equation}
where $q$ is the wavenumber of the excitation. For condensates
with repulsive interactions, the spectrum of long-wavelength
excitations becomes phonon-like, with energies proportional to the
speed of sound waves in the condensate:
\begin {equation}
\label{eq.phononlike} E\approx\hbar c q
\end{equation}
where $c$ is the speed of sound in the condensate. This
modification of the long-wavelength excitation spectrum to
collective behavior is responsible for the phenomenon of
superfluidity (see below).

Condensates with attractive interactions have an imaginary
excitation spectrum and therefore possess an instability;
references are listed in 
Sec.~VI.D.2, below.

\bref

\item
``Collective Excitations of a {B}ose-{E}instein Condensate in a
Dilute Gas,'' D.~S. Jin, J.~R. Ensher, M.~R. Matthews, C.~E.
Wieman, and E.~A. Cornell, Phys. Rev. Lett. {\bf 77}, 420--423
(1996). (I)

\item
``Collective Excitations of a {B}ose-{E}instein Condensate in a
Magnetic Trap,'' M.-O. Mewes, M.~R. Andrews, N.~J. van Druten,
D.~M. Kurn, D.~S. Durfee, C.~G. Townsend, and W.~Ketterle, Phys.
Rev. Lett. {\bf 77}, 988--991 (1996). (I)

\item
``Bose-Einstein Condensates in Time Dependent Traps,'' Y.~Castin
and R.~Dum, Phys. Rev. Lett. {\bf 77}, 5315--5319 (1996). (I)

\item
``Collective Excitations of Atomic {B}ose-{E}instein
Condensates,'' M.~Edwards, P.~A. Ruprecht, K.~Burnett, R.~J. Dodd,
and C.~W. Clark, Phys. Rev. Lett. {\bf 77}, 1671--1674 (1996). (I)

\item
``Collective Excitations of a Trapped {B}ose-Condensed Gas,''
S.~Stringari, Phys. Rev. Lett. {\bf 77}, 2360--2363 (1996). (I)

\item
``Evolution of a Bose Gas in Anisotropic Time-Dependent Traps,''
Yu.~Kagan, E.~L. Surkov, and G.~V. Shylapnikov, Phys. Rev. A {\bf
55}, R18--R21 (1997). (I)

\item
``Propagation of Sound in a {B}ose-{E}instein Condensate,'' M.~R.
Andrews, D.~M. Kurn, H.-J. Miesner, D.~S. Durfee, C.~G. Townsend,
S.~Inouye, and W.~Ketterle, Phys. Rev. Lett. {\bf 79}, 553--556
(1997); {\it ibid.} {\bf{80}}, 2967 (1998). (I)

\item
``Elementary Excitations in Trapped Bose-Einstein Condensed Gases
Beyond the Mean-Field Approximation,'' L.~Pitaevskii and
S.~Stringari, Phys. Rev. Lett. {\bf 81}, 4541--4544 (1998). (I)

\item
``Excitation of Phonons in a {B}ose-{E}instein Condensate by Light
Scattering,'' D.~M. Stamper-Kurn, A.~P. Chikkatur, A.~G{\"o}rlitz,
S.~Inouye, S.~Gupta, D.~E. Pritchard, and W.~Ketterle, Phys. Rev.
Lett. {\bf 83}, 2876--2879 (1999). (I)

\item
``Surface Excitations of a {B}ose-{E}instein Condensate,''
R.~Onofrio, D.~S. Durfee, C.~Raman, M.~K{\"o}hl, C.~E. Kuklewicz,
and W.~Ketterle, Phys. Rev. Lett. {\bf 84}, 810--813 (2000). (I)

\item
``Observation of Harmonic Generation and Nonlinear Coupling in the
Collective Dynamics of a {B}ose-{E}instein Condensate,''
G.~Hechenblaikner, O.~M. Marag{\`o}, E.~Hodby, J.~Arlt,
S.~Hopkins, and C.~J. Foot, Phys. Rev. Lett. {\bf 85}, 692--695
(2000). (I)

\item
``Observation of Quantum Shock Waves Created with Ultra-Compressed
Slow Light Pulses in a {B}ose-{E}instein Condensate,'' Z.~Dutton,
M.~Budde, C.~Slowe, and L.~V. Hau, Science {\bf 293}, 663--668
(2001). (I)

\item
``Experimental Observation of the {B}ogoliubov Transformation for
a {B}ose-{E}instein Condensed Gas,'' J.~M. Vogels, K.~Xu,
C.~Raman, J.~R. Abo-Shaeer, and W.~Ketterle, Phys. Rev. Lett. {\bf
88}, 060402/1--4 (2002). (A)

\item
``Excitation Spectrum of a {B}ose-{E}instein Condensate,''
J.~Steinhauer, R.~Ozeri, N.~Katz, and N.~Davidson, Phys. Rev.
Lett. {\bf 88}, 120407/1--4 (2002). (I)



\eref

\subsubsection{Finite Temperature}

At finite temperatures, there also are thermal excitations as the
condensate interacts with a surrounding cloud of noncondensed
atoms. Two regimes of this interaction have been studied. At low
temperatures and densities, the system is in the collisionless
regime; the mean-field is responsible for the interactions. At
higher temperatures and densities, the system is described by a
pair of coupled hydrodynamic equations within a two-fluid model of
the interactions. The coupling of the condensate to the reservoir
of thermal excitations can both shift the frequencies of the
collective modes and lead to their damping.

\bref


\item
``Dynamics of Trapped Bose Gases at Finite Temperatures,''
E.~Zaremba, T.~Nikuni, and A.~Griffin, J. Low. Temp. Phys. {\bf
116}, 277--347 (1999). Review article. (I)

\item
``{B}ose-{E}instein Condensation in a Dilute Gas: {M}easurement of
Energy and Ground-State Occupation,'' J.~R. Ensher, D.~S. Jin,
M.~R. Matthews, C.~E. Wieman, and E.~A. Cornell, Phys. Rev. Lett.
{\bf 77}, 4984--4987 (1996). (I)

\item
``Temperature-Dependent Damping and Frequency Shifts in Collective
Excitations of a Dilute {B}ose-{E}instein Condensate,'' D.~S. Jin,
M.~R. Matthews, J.~R. Ensher, C.~E. Wieman, and E.~A. Cornell,
Phys. Rev. Lett. {\bf 78}, 764--767 (1997). (I)

\item
``Finite Temperature Excitations of a Trapped {B}ose Gas,''
D.~A.~W. Hutchinson, E.~Zaremba, and A.~Griffin, Phys. Rev. Lett.
{\bf 78}, 1842--1845 (1997). (A)

\item
``Collisionless and Hydrodynamic Excitations of a
{B}ose-{E}instein Condensate,'' D.~M. Stamper-Kurn, H.-J. Miesner,
S.~Inouye, M.~R. Andrews, and W.~Ketterle, Phys. Rev. Lett. {\bf
81}, 500--503 (1998). (I)

\item
``Damping of Low-Energy Excitations of a Trapped Bose-Einstein
Condensate at Finite Temperatures,'' P.~O. Fedichev, G.~V.
Shylapnikov, and J.~T.~M. Walraven, Phys. Rev. Lett. {\bf 80},
2269--2272 (1998). (A)

\item
``Damping in Dilute Bose Gases: A Mean-Field Approach,''
S.~Giorgini, Phys. Rev. A {\bf 57}, 2949--2957 (1998). (A)

\item
``Two-Fluid Hydrodynamics for a Trapped Weakly Interacting Bose
Gas,'' E.~Zaremba, A.~Griffin, and T.~Nikuni, Phys. Rev. A {\bf
57}, 4695--4698 (1998). (A)

\item
``Kinetic Theory of Collective Excitations and Damping in
Bose-Einstein Condensed Gases,'' U. Al Khawaja and H. T. C. Stoof,
Phys. Rev. A {\bf 62}, 053602/1--10 (2000). (A)

\item
``Experimental Observation of {B}eliaev Coupling in a
{B}ose-{E}instein Condensate,'' E.~Hodby, O.~M. Marag{\`o},
G.~Hechenblaikner, and C.~J. Foot, Phys. Rev. Let.. {\bf 86},
2196--2199 (2001). (A)

\item
``Temperature Dependence of Damping and Frequency Shifts of the
Scissors Mode of a Trapped {B}ose-{E}instein Condensate,''
O.~Marag{\`o}, G.~Hechenblaikner, E.~Hodby, and C.~Foot, Phys.
Rev. Lett. {\bf 86}, 3938--3941 (2001). (I)

\item
``Quadrupole Collective Modes in Trapped Finite-Temperature
Bose-Einstein Condensates,'' B. Jackson and E. Zaremba, Phys. Rev.
Lett. {\bf 88}, 180402/1--4 (2002). (A)

\eref

\subsubsection{Solitons}

Solitons are a class of exact solutions to the time-dependent
Gross-Pitaevskii equation in which localized spatial structures
propagate through the condensate medium without changing their
shapes. These spatial structures can be regions of diminished
atomic density (dark solitons) or enhanced atomic density (bright
solitons).

\bref

\item
``Soliton Dynamics in the Collisions of Bose-Einstein Condensates:
an Analogue of the Josephson Effect,'' W.~P. Reinhardt and C.~W.
Clark, J. Phys. B: At. Mol. Opt. Phys. {\bf 30}, L785--L789
(1997). (I)

\item
``Dark Solitons in {B}ose-{E}instein Condensates,'' S.~Burger,
K.~Bongs, S.~Dettmer, W.~Ertmer, K.~Sengstock, A.~Sanpera,
G.~Shlyapnikov, and M.~Lewenstein, Phys. Rev. Lett. {\bf 83},
5198--5201 (1999). (I)

\item
``Generating Solitons by Phase Engineering of a {B}ose-{E}instein
Condensate,'' J.~Denschlag, J.~E. Simsarian, D.~L. Feder, C.~W.
Clark, L.~A. Collins, J.~Cubizolles, L.~Deng, E.~W. Hagley,
K.~Helmerson, W.~P. Reinhardt, S.~L. Rolston, B.~I. Schneider, and
W.~Phillips, Science {\bf 287}, 97--101 (2000). (I)

\item \label{soliton1}
``Formation of a Matter-Wave Bright Soliton,'' L.~Khaykovich,
F.~Schreck, G.~Ferrari, T.~Bourdel, J.~Cubizolles, L.~D. Carr,
Y.~Castin, and C.~Salomon, Science {\bf 296}, 1290--1293 (2002).
(I)

\item \label{soliton2}
``Formation and Propagation of Matter-Wave Soliton Trains,'' K.~E.
Strecker, G.~B. Partridge, A.~G. Truscott, and R.~G. Hulet, Nature
{\bf 417}, 150--153 (2002). (I)

\item
``Dynamics of Dark Solitons in Elongated Bose-Einstein
Condensates,'' A.~Muryshev, G.~V. Shlyapnikov, W.~Ertmer,
K.~Sengstock, and M.~Lewenstein, Phys. Rev. Lett. {\bf 89},
110401/1--4 (2002). (I)

\eref

\subsubsection{Superfluidity}

The crossover between the low-energy excitations and high-energy
excitations mentioned above is the basis for the existence of
superfluidity, or dissipationless mass flow, in a Bose-Einstein
condensate. One criterion of superfluidity is that obstacles
moving within the condensate must exceed a critical velocity
before they can transfer momentum to the condensate. The Landau
criterion for superfluidity states:
\begin{equation}
v_\mathrm{critical} = \min \left(\frac{E_p}{p}\right)
\end{equation}
where $E_p$ and $p$ are the energy and momentum of an excitation,
and $v_\mathrm{critical}$ is the velocity below which no
excitations can be generated and superfluid flow is possible.

For an excitation spectrum that is particle-like,
Eq.~\ref{eq.particlelike}, it is clear that the critical velocity
is zero, that is, no superfluid flow is possible. For a
phonon-like excitation spectrum, Eq.~\ref{eq.phononlike}, however,
the critical velocity is the speed of sound, and superfluidity
becomes possible.

\bref


\item
``Observation of the Scissors Mode and Evidence for Superfluidity
of a Trapped {B}ose-{E}instein Condensed Gas,'' O.~M. Marag{\'o},
S.~A. Hopkins, J.~Arlt, E.~Hodby, G.~Heckenblaikner, and C.~J.
Foot, Phys. Rev. Lett. {\bf 84}, 2056--2059 (2000). (I)

\item
``Suppression and Enhancement of Impurity Scattering in a
{B}ose-{E}instein Condensate,'' A.~P. Chikkatur, A.~G{\"o}rlitz,
D.~M. Stamper-Kurn, S.~Inouye, S.~Gupta, and W.~Ketterle, Phys.
Rev. Lett. {\bf 85}, 483--486 (2000). (I)

\item
``Observation of Superfluid Flow in a {B}ose-{E}instein Condensed
Gas,'' R.~Onofrio, C.~Raman, J.~M. Vogels, J.~R. Abo-Shaeer, A.~P.
Chikkatur, and W.~Ketterle, Phys. Rev. Lett. {\bf 85}, 2228--2231
(2000). (I)

\item
``Superfluid and Dissipative Dynamics of a {B}ose-{E}instein
Condensate in a Periodic Optical Potential,'' S.~Burger, F.~S.
Cataliotti, C.~Fort, F.~Minardi, M.~Inguscio, M.~L. Chiofalo, and
M.~P. Tosi, Phys. Rev. Lett. {\bf 86}, 4447--4450 (2001). (I) 

\item
``Direct Observation of Irrotational Flow and Evidence of
Superfluidity in a Rotating {B}ose-{E}instein Condensate,''
G.~Hechenblaikner, E.~Hodby, S.~A. Hopkins, O.~M. Marag{\`o}, and
C.~J. Foot, Phys. Rev. Lett. {\bf 88}, 070406/1--4 (2002). (I)

\eref

\subsubsection{Vortices}

According to the hydrodynamic formulation of the GP equation, the
gradient of the phase of the condensate order parameter is related
to the condensate velocity by
\begin{equation}
\mathbf{v} = \frac{\hbar}{m}\nabla \phi.
\end{equation}
The phase is therefore a ``potential'' for the velocity field of
the condensate, and thus does not permit rotational flow in a
simply-connected geometry (since $\nabla \times \mathbf{v}=0$
always). In a multiply-connected geometry, however, in which the
condensate density goes to zero in some region (such as a toroidal
trap), the only restriction on the order parameter is that it be
single-valued. The phase variation in any loop traced around the
excluded region is restricted to $2n\pi$, where $n$ is any
integer. The corresponding mass flow is a quantized vortex of
winding number $n$.

Vortices were one of the original signs of superfluidity in liquid
helium, but their generation and detection in dilute-gas
condensates eluded experimenters for several years. This initial
frustration has led since to a recent blossoming of experimental
techniques and realizations of vortices --- from single vortices
in multiple-spin state condensates to lattices of vortices within
a single condensate. A sampling of the relevant literature is
given below.

\bref

\item
``Vortices in a Trapped Dilute {B}ose-{E}instein Condensate,''
A.~L. Fetter and A.~A. Svidzinsky, J. Phys.: Condens. Matter {\bf
13}, R135--R194 (2001). Review article. (I)

\item
``Vortex Stability and Persistent Currents in Trapped Bose
Gases,'' D.~S. Rokhsar, Phys. Rev. Lett. {\bf 79}, 2164--2167
(1997). (I)

\item
``Excitation Spectroscopy of Vortex States in Dilute Bose-Einstein
Condensed Gases,'' R.~J. Dodd, K.~Burnett, M.~Edwards, and C.~W.
Clark, Phys. Rev. A {\bf 56}, 587--590 (1997). (I)

\item
``Predicted Signatures of Rotating {B}ose-{E}instein
Condensates,'' D.~A. Butts and D.~S. Rokhsar, Nature {\bf 397},
327--329 (1999). (I)

\item
``Phase Diagram of Quantized Vortices in a Trapped Bose-Einstein
Condensed Gas,'' S.~Stringari, Phys. Rev. Lett. {\bf 82},
4371--4375 (1999). (I)

\item
``Vortex Stability of Interacting Bose-Einstein Condensates
Confined in Anisotropic Harmonic Traps,'' D.~L. Feder, C.~W.
Clark, and B.~I. Schneider, Phys. Rev. Lett. {\bf 82}, 4956--4959
(1999). (I)

\item
\label{Matthews1999} ``Vortices in a {B}ose-{E}instein
Condensate,'' M.~R. Matthews, B.~P. Anderson, P.~C. Haljan, D.~S.
Hall, C.~E. Wieman, and E.~A. Cornell, Phys. Rev. Lett. {\bf 83},
2498--2501 (1999). (I)

\item
``Vortex Formation in a Stirred {B}ose-{E}instein Condensate,''
K.~W. Madison, F.~Chevy, W.~Wohlleben, and J.~Dalibard, Phys. Rev.
Lett. {\bf 84}, 806--809 (2000). (I)

\item
``Measurement of the Angular Momentum of a Rotating
{B}ose-{E}instein Condensate,'' F.~Chevy, K.~W. Madison, and
J.~Dalibard, Phys. Rev. Lett. {\bf 85}, 2223--2227 (2000). (I)

\item
``Vortex Precession in {B}ose-{E}instein Condensates:
{O}bservations with Filled and Empty Cores,'' B.~P. Anderson,
P.~C. Haljan, C.~E. Wieman, and E.~A. Cornell, Phys. Rev. Lett.
{\bf 85}, 2857--2860 (2000). (I)

\item
``Use of Surface-Wave Spectroscopy to Characterize Tilt Modes of a
Vortex in a {B}ose-{E}instein Condensate,'' P.~C. Haljan, B.~P.
Anderson, I.~Coddington, and E.~A. Cornell, Phys. Rev. Lett. {\bf
86}, 2922--2925 (2001). (A)

\item
``Watching Dark Solitons Decay into Vortex Rings in a
{B}ose-{E}instein Condensate,'' B.~P. Anderson, P.~C. Haljan,
C.~A. Regal, D.~L. Feder, L.~A. Collins, C.~W. Clark, and E.~A.
Cornell, Phys. Rev. Lett. {\bf 86}, 2926--2929 (2001). (I)

\item
``Stationary States of a Rotating {B}ose-{E}instein Condensate:
{R}outes to Vortex Nucleation,'' K.~W. Madison, F.~Chevy,
V.~Bretin, and J.~Dalibard, Phys. Rev. Lett. {\bf 86}, 4443--4446
(2001). (I)

\item
``Observation of Vortex Phase Singularities in {B}ose-{E}instein
Condensates,'' S.~Inouye, S.~Gupta, T.~Rosenband, A.~P. Chikkatur,
A.~G{\"o}rlitz, T.~L. Gustavson, A.~E. Leanhardt, D.~E. Pritchard,
and W.~Ketterle, Phys. Rev. Lett. {\bf 87}, 080402/1--4 (2001).
(I)

\item
``Vortex Nucleation in a Stirred {B}ose-{E}instein Condensate,''
C.~Raman, J.~R. Abo-Shaeer, J.~M. Vogels, K.~Xu, and W.~Ketterle,
Phys. Rev. Lett. {\bf 87}, 210402/1--4 (2001). (I)

\item
``Driving {B}ose-{E}instein Condensate Vorticity with a Rotating
Normal Cloud,'' P.~C. Haljan, I.~Coddington, P.~Engels, and E.~A.
Cornell, Phys. Rev. Lett. {\bf 87}, 210403/1--4 (2001). (I)

\item
``Observation of Vortex Lattices in {B}ose-{E}instein
Condensates,'' J.~R. Abo-Shaeer, C.~Raman, J.~M. Vogels, and
W.~Ketterle, Science {\bf 292}, 476--479 (2001). (I)

\item
``Vortex Nucleation in {B}ose-{E}instein Condensates in an Oblate,
Purely Magnetic Potential,'' E.~Hodby, G.~Heckenblaikner, S.~A.
Hopkins, O.~M. Marag{\'o}, and C.~J. Foot, Phys. Rev. Lett. {\bf
88}, 010405/1--4 (2002). (I)

\item
``Formation and Decay of Vortex Lattices in {B}ose-{E}instein
Condensates at Finite Temperatures,'' J.~R. Abo-Shaeer, C.~Raman,
and W.~Ketterle, Phys. Rev. Lett. {\bf 88}, 070409/1--4 (2002).
(I)


\eref

\subsubsection{Formation}

The process by which a Bose-Einstein condensate forms out of a
thermal gas as it is cooled below the transition temperature is
becoming increasingly well understood. It has been known for some
time that there is an enhancement of scattering into the
condensate proportional to the number of atoms already in it
(Bosonic stimulation) and that the process of condensate formation
builds up a long-range coherence among its constituent atoms in a
finite amount of time. A recent quantum-kinetic-theory approach to
condensate formation finds rather good agreement with the
available experiments, although some features of the data have not
been yet fully explained.

Condensates also have been formed out of a cold thermal gas by
modifying the trapping potential, thereby increasing the
phase-space density of the atomic cloud. The process is adiabatic
and therefore reversible, and provides another viewpoint on the
formation process.

\bref

\item
``Formation of the Condensate in a Dilute {B}ose Gas,'' H.~T.~C.
Stoof, Phys. Rev. Lett. {\bf 66}, 3148--3151 (1991). (A)

\item
``Initial Stages of Bose-Einstein Condensation,'' H.~T.~C. Stoof,
Phys. Rev. Lett. {\bf 78}, 768--771 (1997). (I)

\item
``Kinetics of {B}ose-{E}instein Condensation in a Trap,'' C.~W.
Gardiner, P.~Zoller, R.~J. Ballagh, and M.~Davis, Phys. Rev. Lett.
{\bf 79}, 1793--1796 (1997). (I)

\item
``Bosonic Stimulation in the Formation of a {B}ose-{E}instein
Condensate,'' H.-J. Miesner, D.~M. Stamper-Kurn, M.~R. Andrews,
D.~S. Durfee, S.~Inouye, and W.~Ketterle, Science {\bf 279},
1005--1007 (1998). (I)

\item
``Reversible Formation of a {B}ose-{E}instein Condensate,'' D.~M.
Stamper-Kurn, H.-J. Miesner, A.~P. Chikkatur, S.~Inouye,
J.~Stenger, and W.~Ketterle, Phys. Rev. Lett. {\bf 81}, 2194--2197
(1998). (I)

\item
``Quantum Kinetic Theory. VI. The Growth of a {B}ose-{E}instein
Condensate,'' M.~D. Lee and C.~W. Gardiner, Phys. Rev. A {\bf 62},
033606/1--26 (2000); and references therein. (A)

\item
``Direct Observation of Growth and Collapse of a Bose-Einstein
Condensate with Attractive Interactions,'' J.~M. Gerton,
D.~Strekalov, I.~Prodan, and R.~G. Hulet, Nature {\bf 408},
692--695 (2000). (I)

\item
``Growth of {B}ose-{E}instein Condensates from Thermal Vapor,''
M.~K{\"o}hl, M.~J. Davis, C.~W. Gardiner, T.~W. H{\"a}nsch, and
T.~Esslinger, Phys. Rev. Lett. {\bf 88}, 080402/1--4 (2002). (I)

\eref

\subsection{Effects of Interatomic Interactions}

As mentioned above, nearly every aspect of the condensate depends
critically on the $s$-wave scattering length $a$. From
measurements of macroscopic properties of the condensate, such as
size and shape, one can gain insight into binary elastic
collisions. Two and three-body inelastic collisions also play an
important role in the fate of the condensate. The following review
articles present some of the details of ultracold collisions that
are relevant to Bose-Einstein condensates.

Atoms involved in elastic collisions (``good'' collisions) retain
their internal states but can redistribute momentum and energy.
These collisions are responsible for the thermalization of the
samples that permits evaporative cooling to proceed efficiently.
Inelastic collisions (``bad'' collisions) often result in heating
and trap loss, although they are sometimes subjects of interest in
their own right. See also Ref.~[\ref{evap}] above, and
Ref.~[\ref{Burt1997}] below.

\bref

\item
``Experiments and Theory in Cold and Ultracold Collisions,'' J.
Weiner, V.~S. Bagnato, S. Zilio, and P.~S. Julienne, Rev. Mod.
Phys. {\bf 71}, 1--85 (1999). Review article. (I)

\item
``Quantum Encounters of the Cold Kind,'' K.~Burnett, P.~S.
Julienne, P.~D. Lett, E.~Tiesinga, and C.~J. Williams, Nature {\bf
416}, 225--232 (2002). Review article. (E)

\item
``Dipolar Decay in Two Recent Bose-Einstein Condensation
Experiments,'' H.~M.~J.~M. Boesten, A.~Moerdijk, and B.~J.
Verhaar, Phys. Rev. A {\bf 54}, R29--R32 (1996). (I)

\item ``Avalanches in a {B}ose-{E}instein Condensate,''
J.~Schuster, A.~Marte, S.~Amtage, B.~Sang, G.~Rempe, and H.~C.~W.
Beijerinck, Phys. Rev. Lett. {\bf 87}, 170404/1--4 (2001). (A)

\eref

\subsubsection{Tunable Interactions: Feshbach Resonances}
\label{Feshbach}

In certain instances the $s$-wave scattering length $a$ can depend
upon the value of a parameter external to the condensate, such as
the magnetic field. Loosely speaking, two atoms can tunnel into a
bound state in which the binding energy is stored internally by
flipping the spin of one of the atoms. The bound-state energies
can depend on the value of the magnetic field, which means that
the tunnelling (and hence $a$, by the scattering cross-section)
can be enhanced or reduced by simply varying the magnetic field.
This is the Feshbach resonance.

Experiments have found that inelastic processes also tend to
increase in the vicinity of a Feshbach resonance, in some cases to
the point at which experiments to tune $a$ are rendered
impossible. Stable condensates with tunable interactions in
${}^{85}$Rb~[\ref{rubidium}] and ${}^{133}$Cs~[\ref{cesium}]
nevertheless have been created to date. Another use of Feshbach
resonances has been to create bright solitons (see
Refs.~[\ref{soliton1}] and~[\ref{soliton2}]).

\bref

\item
``Feshbach Resonances in Atomic Bose-Einstein Condensates,'' E.
Timmermans, P. Tommasini, M. Hussein, and A. Kerman, Phys. Rep.
{\bf 315}, 199--230 (1999). Review article. (I)

\item
``Threshold and Resonance Phenomena in Ultracold Ground-State
Collisions,'' E.~Tiesinga, B.~J. Verhaar, and H.~T.~C. Stoof,
Phys. Rev. A {\bf 47}, 4114--4122 (1993). (A)

\item
``Influence of Nearly Resonant Light on the Scattering Length in
Low-Temperature Atomic Gases,'' P.~O. Fedichev, Yu.~Kagan, G.~V.
Shlyapnikov, and J.~T.~M. Walraven, Phys. Rev. Lett. {\bf 77},
2913--2916 (1996). (A)

\item
``Observation of {F}eshbach Resonances in a {B}ose-{E}instein
Condensate,'' S.~Inouye, M.~R. Andrews, J.~Stenger, H.-J. Miesner,
D.~M. Stamper-Kurn, and W.~Ketterle, Nature {\bf 392}, 151--154
(1998). (I)

\item
``Controlling Atom-Atom Interaction at Ultralow Temperatures by dc
Electric Fields,'' M.~Marinescu and L.~You, Phys. Rev. Lett. {\bf
81}, 4596--4599 (1998). (A)

\item
``Strongly Enhanced Inelastic Collisions in a {B}ose-{E}instein
Condensate near {F}eshbach Resonances,'' J.~Stenger, S.~Inouye,
M.~R. Andrews, H.-J. Miesner, D.~M. Stamper-Kurn, and W.~Ketterle,
Phys. Rev. Lett. {\bf 82}, 2422--2425 (1999). (I)

\item
``Time-Dependent Feshbach Resonance Scattering and Anomalous Decay
of a Na Bose-Einstein Condensate,'' F.~A. van~Abeelen and B.~J.
Verhaar, Phys. Rev. Lett. {\bf 83}, 1550--1553 (1999). (I)

\item
``Microscopic Dynamics in a Strongly Interacting {B}ose-{E}instein
Condensate,'' N.~R. Claussen, E.~A. Donley, S.~T. Thompson, and
C.~E. Wieman, Phys. Rev. Lett. {\bf 89}, 010401/1--4 (2002). (A)

\item
``Mean-Field Theory of Feshbach-Resonant Interactions in
${}^{85}$Rb Condensates,'' M.~Mackie, K.-A. Suominen, and
J.~Javanainen, Phys. Rev. Lett. {\bf 89}, 180403/1--4 (2002). (A)


\eref

\subsubsection{Negative Scattering Length}
\label{attractive}

If the scattering length is negative, the interatomic interactions
are attractive. In a homogeneous gas the excitation frequencies
are imaginary, which corresponds to an instability that prevents
condensates from forming. In an inhomogenous gas, the kinetic
energy can help stabilize against collapse, although for
sufficiently large numbers of atoms the attractive interactions
cause the condensate to collapse.

Recent experiments also have induced a collapse as the scattering
length is switched rapidly from positive to negative using a
Feshbach resonance (see above). The condensate rapidly implodes,
generating subsequent dynamics that often exhibit explosions of
atoms (the ``Bosenova'') and other behavior that is not yet fully
understood.

\bref

\item
``Role of Attractive Interactions on Bose-Einstein Condensation,''
R.~J. Dodd, M.~Edwards, C.~J. Williams, C.~W. Clark, M.~J.
Holland, P.~A. Ruprecht, and K.~Burnett, Phys. Rev. A {\bf 54},
661--664 (1996). (I)

\item
``Evolution and Global Collapse of Trapped Bose Condensates under
Variations of the Scattering Length,'' Yu.~Kagan, E.~L. Surkov,
and G.~V. Shlyapnikov, Phys. Rev. Lett. {\bf 79}, 2604--2607
(1997). (I)

\item
``Macroscopic Quantum Tunneling of a Bose-Einstein Condensate with
Attractive Interaction,'' M.~Ueda and A.~J. Leggett, Phys. Rev.
Lett. {\bf 80}, 1576--1579 (1998). (A)

\item
``Measurements of Collective Collapse in a {B}ose-{E}instein
Condensate with Attractive Interactions,'' C.~A. Sackett, J.~M.
Gerton, M.~Welling, and R.~G. Hulet, Phys. Rev. Lett. {\bf 82},
876--879 (1999). (I)

\item
``Controlled Collapse of a {B}ose-{E}instein Condensate,'' J.~L.
Roberts, N.~R. Claussen, S.~L. Cornish, E.~A. Donley, E.~A.
Cornell, and C.~E. Wieman, Phys. Rev. Lett. {\bf 86}, 4211--4214
(2001). (I)

\item
\label{Donley2001} ``Dynamics of Collapsing and Exploding
{B}ose-{E}instein Condensates,'' E.~A. Donley, N.~R. Claussen,
S.~L. Cornish, J.~L. Roberts, E.~A. Cornell, and C.~E. Wieman,
Nature {\bf 412}, 295--299 (2001). (I)

\item
``Intermittent Implosion and Pattern Formation of Trapped
Bose-Einstein Condensates with an Attractive Interaction,''
H.~Saito and M.~Ueda, Phys. Rev. Lett. {\bf 86}, 1406--1409
(2001). (A)

\eref

\subsection{Condensates as Matter Waves}

\subsubsection{Phase Coherence}

\normalsize

The description of a Bose-Einstein condensate in terms of an order
parameter carries with it the notion of a phase; indeed, we have
already seen that the phase plays the role of a velocity-field
potential in the hydrodynamic description of the BEC. The phase,
by itself, is not observable. If two condensates are brought
together, however, interference patterns are expected to emerge
based on the relative phase between the two condensates.

Complicating matters slightly, the number-phase uncertainty
relation,
\begin{equation}
\Delta N \Delta \phi \approx 1,
\end{equation}
prevents complete knowledge of both the number of atoms in the
condensate and its phase. A condensate in a number state has a
vanishing order parameter and thus cannot be said to have a
well-defined phase at all. Nevertheless, interference patterns
still emerge as the detection process entangles the two
condensates.

These papers examine interference between two (or more)
condensates, both in space (where the interference manifests
itself in the atomic density) and in time (using different
internal states).

\bref

\item
``Quantum Phase of a {B}ose-{E}instein Condensate with an
Arbitrary Number of Atoms,'' J.~Javanainen and S.~M. Yoo, Phys.
Rev. Lett. {\bf 76}, 161--164 (1996). (I)

\item \label{andrews}
``Observation of Interference between Two {B}ose-{E}instein
Condensates,'' M.~R. Andrews, C.~G. Townsend, H.-J. Miesner, D.~S.
Durfee, D.~M. Kurn, and W.~Ketterle, Science {\bf 275}, 637--641
(1997). (I)

\item
``Transition from Phase Locking to the Interference of Independent
{B}ose Condensates: Theory versus Experiment,'' A.~R{\"o}hrl,
M.~Naraschewski, A.~Schenzle, and H.~Wallis, Phys. Rev. Lett. {\bf
78}, 4143--4146 (1997). (A)

\item
``Measurements of Relative Phase in Two-Component
{B}ose-{E}instein Condensates,'' D.~S. Hall, M.~R. Matthews, C.~E.
Wieman, and E.~A. Cornell, Phys. Rev. Lett. {\bf 81}, 1543--1546
(1998). (I)

\item
``Phase Standard for Bose-Einstein Condensates,'' J.~A. Dunningham
and K.~Burnett, Phys. Rev. Lett. {\bf 82}, 3729--3733 (1999). (I)

\item
``Measurement of the Coherence of a {B}ose-{E}instein
Condensate,'' E.~Hagley, L.~Deng, M.~Kozuma, M.~Trippenbach,
Y.~Band, M.~Edwards, M.~Doery, P.~Julienne, K.~Helmerson,
S.~Rolston, and W.~Phillips, Phys. Rev. Lett. {\bf 83}, 3112--3115
(1999). (I)

\item
``Imaging the Phase of an Evolving {B}ose-{E}instein Condensate
Wave Function,'' J.~E. Simsarian, J.~Denschlag, M.~Edwards, C.~W.
Clark, L.~Deng, E.~W. Hagley, K.~Helmerson, S.~L. Rolston, and
W.~D. Phillips, Phys. Rev. Lett. {\bf 85}, 2040--2043 (2000). (I)

\item
``Measurement of the Spatial Coherence of a Trapped {B}ose Gas at
the Phase Transition,'' I.~Bloch, T.~W. H{\"a}nsch, and
T.~Esslinger, Nature {\bf 403}, 166--170 (2000). (I)

\item
``Exploring Phase Coherence in a 2{D} Lattice of {B}ose-{E}instein
Condensates,'' M.~Greiner, I.~Bloch, O.~Mandel, T.~W. H{\"a}nsch,
and T.~Esslinger, Phys. Rev. Lett. {\bf 87}, 160405/1--4
(2001). (I) 

\item
``Observation of Phase Fluctuations in Elongated {B}ose-{E}instein
Condensates,'' S.~Dettmer, D.~Hellweg, P.~Ryytty, J.~J. Arlt,
W.~Ertmer, K.~Sengstock, D.~S. Petrov, G.~V. Shlyapnikov,
H.~Kreutzmann, L.~Santos, and M.~Lewenstein, Phys. Rev. Lett. {\bf
87}, 160406/1--4 (2001). (A)

\item
``Time-Domain Atom Interferometry across the Threshold for
{B}ose-{E}instein Condensation,'' F.~Minardi, C.~Fort,
P.~Maddaloni, M.~Modugno, and M.~Inguscio, Phys. Rev. Lett. {\bf
87}, 170401/1--4 (2001). (I)

\item
``Expansion of a Coherent Array of {B}ose-{E}instein
Condensates,'' P.~Pedri, L.~Pitaevskii, S.~Stringari, C.~Fort,
S.~Burger, F.~S. Cataliotti, P.~Maddaloni, F.~Minardi, and
M.~Inguscio, Phys. Rev. lett. {\bf 87}, 220401/1--4 (2001). (I)

\item
``Dynamics of a Bose-Einstein Condensate at Finite Temperature in
an Atom-Optical Coherence Filter,'' F.~Ferlaino, P.~Maddaloni,
S.~Burger, F.~S. Cataliotti, C.~Fort, M.~Modugno, and M.~Inguscio,
Phys. Rev. A {\bf 66}, 011604/1--4 (2002). (I)
 \eref

\subsubsection{Higher-Order Coherence}

Bose-Einstein condensates also are expected to have perfect
higher-order coherence as well as first-order phase coherence.
These coherences manifest themselves in the two and three-body
correlations, which are discovered experimentally by examining the
mean-field energy and the three-body recombination rate,
respectively.

\bref

\item
\label{Burt1997} ``Coherence, Correlations, and Collisions: {W}hat
One Learns about {B}ose-{E}instein Condensates from Their Decay,''
E.~A. Burt, R.~W. Ghrist, C.~J. Myatt, M.~J. Holland, E.~A.
Cornell, and C.~E. Wieman, Phys. Rev. Lett. {\bf 79}, 337--340
(1997). (I)

\item
``Coherence Properties of {B}ose-{E}instein Condensates and Atom
Lasers,'' W.~Ketterle and H.-J. Miesner, Phys. Rev. A {\bf 56},
3291--3293 (1997). (I) 

\item
``Characterizing the Coherence of Bose-Einstein Condensates and
Atom Lasers,'' R.~Dodd, C.~W. Clark, M.~Edwards, and K.~Burnett,
Opt. Expr. {\bf 1}, 284--292 (1997). (I) 

\item
``Spatial Coherence and Density Correlations of Trapped Bose
Gases,'' M.~Naraschewski and R.~J. Glauber, Phys. Rev. A {\bf 59},
4595--4607 (1999). (A)

\eref

\subsubsection{Tunnelling, Bloch Oscillations, and Josephson Effect}

A lattice of Bose-Einstein condensates can be produced by placing
a single condensate into a series of microtraps formed by an
optical standing wave. Typically, the condensate atoms can tunnel
from site to site within the lattice, exhibiting collective
behavior reminiscent of the Josephson effect in superconductors.
They maintain a phase coherence between the lattice sites by
occupying delocalized states. When the trap depths are increased,
however, the tunnelling becomes suppressed, and the lattice sites
contain number-squeezed states --- that is, the phase coherence
between sites is lost, and well-defined numbers of atoms occupy
each lattice site.

\bref

\item
``Boson Localization and the Superfluid-Insulator Transition,''
M.~P.~A. Fisher, P.~B. Weichman, G.~Grinstein, and D.~S. Fisher,
Phys. Rev. B {\bf 40}, 546--570 (1989). (I)

\item
``Cold Bosonic Atoms in Optical Lattices,'' D. Jaksch, C. Bruder,
J.~I. Cirac, C.~W. Gardiner, and P. Zoller, Phys. Rev. Lett. {\bf
81}, 3108--3111 (1998). (I)

\item \label{brian}
``Macroscopic Quantum Interference from Atomic Tunnel Arrays,''
B.~P. Anderson and M.~A. Kasevich, Science {\bf 282}, 1686--1689
(1998). (I)

\item
``Squeezed States in a {B}ose-{E}instein Condensate,'' C.~Orzel,
A.~K. Tuchman, M.~L. Fenselau, M.~Yasuda, and M.~A. Kasevich,
Science {\bf 291}, 2386--2389 (2001). (I) 

\item
``Dynamical Tunnelling of Ultracold Atoms,'' W.~K. Hensinger,
H.~H{\"a}ffner, A.~Browaeys, N.~R. Heckenberg, K.~Helmerson,
C.~McKenzie, G.~J. Milburn, W.~D. Phillips, S.~L. Rolston,
H.~Rubinsztein-Dunlop, and B.~Upcroft, Nature {\bf 412}, 52--55
(2001). (I) 

\item
``Josephson Junction Arrays with {B}ose-{E}instein Condensates,''
F.~S. Cataliotti, S.~Burger, C.~Fort, P.~Maddaloni, F.~Minardi,
A.~Trombettoni, A.~Smerzi, and M.~Inguscio, Science {\bf 293},
843--846 (2001). (I) 

\item
``{B}loch Oscillations and Mean-Field Effects of {B}ose-{E}instein
Condensates in 1{D} Optical Lattices,'' O.~Morsch, J.~H.
M{\"u}ller, M.~Cristiani, D.~Ciampini, and E.~Arimondo, Phys. Rev.
Lett. {\bf 87}, 140402/1--4 (2001). (I) 

\item
``Quantum Phase Transition from a Superfluid to a {M}ott Insulator
in a Gas of Ultracold Atoms,'' M.~Greiner, O.~Mandel,
T.~Esslinger, T.~W. H{\"a}nsch, and I.~Bloch, Nature {\bf 415},
39--44 (2002). (I)

\eref

\subsubsection{Collapses and Revivals}

A condensate wave packet can be expanded in states of definite
number, $N$, each with a related chemical potential $\mu_N$. Each
of these states has a phase that evolves in time as $N\mu_N
t/\hbar$. Calculation of the macroscopic wavefunction in terms of
these states for relatively small numbers of atoms in the
condensate reveals a periodic collapse and revival of the
condensate phase. These collapses and revivals affect the
interference properties of the condensate. In the thermodynamic
limit ($\overline{N} \rightarrow \infty$) the collapse time
becomes infinitely long and the condensate phase is well-defined.
It is in this limit that the macroscopic wavefunction can be
considered an order parameter.

\bref

\item
``Quantum Phase Diffusion of a Bose-Einstein Condensate,''
M.~Lewenstein and L.~You, Phys. Rev. Lett. {\bf 77}, 3489--3493
(1996). (A)

\item
``Collapses and Revivals in the Interference between Two
Bose-Einstein Condensates Formed in Small Atomic Samples,'' E.~M.
Wright, T.~Wong, M.~J. Collett, S.~M. Tan, and D.~F. Walls, Phys.
Rev. A {\bf 56}, 591--602 (1997). (I)

\item
``Collapse and Revival of the Matter Wave Field of a Bose-Einstein
Condensate,'' M.~Greiner, O. Mandel, T.~W. H{\"a}nsch, and
I.~Bloch, Nature {\bf 419}, 51--54 (2002). (I)

\eref

\subsection{Condensate Optics}

\subsubsection{Light and Matter Waves}

The interaction between coherent light and coherent matter waves
produces some of the most fascinating behavior involving
Bose-Einstein condensates. Bragg diffraction involves the coherent
scattering of light from one laser beam into another, with a
corresponding atomic recoil that conserves energy and momentum. In
superradiant Rayleigh scattering, the initial incoherent
(spontaneous) scattering of light by a condensate establishes a
moving matter-wave grating as a fraction of the condensate recoils
to conserve momentum. This induced grating then reinforces itself
by further stimulated scattering of the incident light beam such
that additional atoms join the single moving mode. The matter-wave
amplification occurring in superradiant scattering has also been
initialized with a controlled``seed'' condensate that forms a
moving matter-wave grating in conjunction with a larger condensate
serving as a ``gain medium.'' The stimulated scattering in this
case reinforces the initially seeded mode.

Coherent matter waves can also interact with each other in a
process called four-wave mixing. In this nonlinear effect, three
condensates with different wave vectors $\mathbf{k_1}$,
$\mathbf{k_2}$, and $\mathbf{k_3}$ interact to generate a fourth
condensate $\mathbf{k_4}=\mathbf{k_1}-\mathbf{k_2}+\mathbf{k_3}$.

\bref

\item
``Nonlinear and Quantum Atom Optics,'' S.~L. Rolston and W.~D.
Phillips, Nature {\bf 416}, 219--224 (2002). Review article. (E)

\item
``Four Wave Mixing in the Scattering of Bose-Einstein
Condensates,'' M.~Trippenbach, Y.~B. Band, and P.~S. Julienne,
Opt. Expr. {\bf 3}, 530--537 (1998). (I)

\item
``Amplifying an Atomic Wave Signal using a Bose-Einstein
Condensate,'' C.~K. Law and N.~P. Bigelow, Phys. Rev. A {\bf 58},
4791--4795 (1998). (A)

\item
``Coherent Splitting of {B}ose-{E}instein Condensed Atoms with
Optically Induced {B}ragg Diffraction,'' M.~Kozuma, L.~Deng, E.~W.
Hagley, J.~Wen, R.~Lutwak, K.~Helmerson, S.~L. Rolston, and W.~D.
Phillips, Phys. Rev. Lett. {\bf 82}, 871--875 (1999). (I)

\item
``Diffraction of a Released {B}ose-{E}instein Condensate by a
Pulsed Standing Light Wave,'' Y.~B. Ovchinnikov, J.~H. M{\"u}ller,
M.~R. Doery, E.~J.~D. Vredenbregt, K.~Helmerson, S.~L. Rolston,
and W.~D. Phillips, Phys. Rev. Lett. {\bf 83}, 284--287 (1999).
(I)

\item
``Temporal, Matter-Wave-Dispersion {T}albot Effect,'' L.~Deng,
E.~W. Hagley, J.~Denschlag, J.~E. Simsarian, M.~Edwards, C.~W.
Clark, K.~Helmerson, S.~L. Rolston, and W.~D. Phillips, Phys. Rev.
Lett. {\bf 83}, 5407--5411 (1999). (I)

\item
``Superradiant {R}ayleigh Scattering from a {B}ose-{E}instein
Condensate,'' S.~Inouye, A.~P. Chikkatur, D.~M. Stamper-Kurn,
J.~Stenger, D.~E. Pritchard, and W.~Ketterle, Science {\bf 285},
571--574 (1999). (I)

\item
``Four-Wave Mixing with Matter Waves,'' L.~Deng, E.~W. Hagley,
J.~Wen, M.~Trippenbach, Y.~Band, P.~S. Julienne, J.~Simsarian,
K.~Helmerson, S.~L. Rolston, and W.~D. Phillips, Nature {\bf 398},
218--220 (1999). (I)

\item
``Phase-Coherent Amplification of Matter Waves,''M.~Kozuma,
Y.~Suzuki, Y.~Torii, T.~Sugiura, T.~Kuga, E.~W. Hagley, and
L.~Deng, Science {\bf 286}, 2309--2312 (1999). (I)

\item
``Phase-Coherent Amplification of Atomic Matter Waves,''
S.~Inouye, T.~Pfau, S.~Gupta, A.~P. Chikkatur, A.~G{\"o}rlitz,
D.~E. Pritchard, and W.~Ketterle, Nature {\bf 402}, 641--644
(1999). (I)

\item
``Theory of Superradiant Scattering of Laser Light from
Bose-Einstein Condensates,'' M.~G. Moore and P.~Meystre, Phys.
Rev. Lett. {\bf 83}, 5202--5205 (1999). (A)


\item
``Amplification of Light and Atoms in a {B}ose-{E}instein
Condensate,'' S.~Inouye, R.~F. L{\"o}w, S.~Gupta, T.~Pfau,
A.~G{\"o}rlitz, T.~L. Gustavson, D.~E. Pritchard, and W.~Ketterle,
Phys. Rev. Lett. {\bf 85}, 4225--4228 (2000). (I)

\eref

\subsubsection{Atom Lasers}

An atom laser is generated by coherently removing atoms from a
Bose-Einstein condensate, typically by interaction with applied
laser light or a radiofrequency electromagnetic field. Both pulsed
and quasicontinuous atom-laser beams have been demonstrated. (See
the preceding section for the demonstration of a coherent gain
mechanism.) A mode-locked atom laser has also been demonstrated
(see Ref.[\ref{brian}]).

\bref

\item
``A Beginner's Guide to the Atom Laser,'' D. Kleppner, Phys. Today
{\bf 50}(8), 11--13 (August 1997). (E)

\item
``Atom Lasers,'' K. Helmerson, D. Hutchinson, K. Burnett, and
W.~D. Phillips, Phys. World {\bf 12} (8), 31--35 (1999). Popular
review article. (E)

\item
``Output Coupler for {B}ose-{E}instein Condensed Atoms,'' M.-O.
Mewes, M.~R. Andrews, D.~M. Kurn, D.~S. Durfee, C.~G. Townsend,
and W.~Ketterle, Phys. Rev. Lett. {\bf 78}, 582--585 (1997). (I)

\item
``Theory of an Output Coupler for {B}ose-{E}instein Condensed
Atoms,'' R.~J. Ballagh, K.~Burnett, and T.~F. Scott, Phys. Rev.
Lett. {\bf 78}, 1607--1611 (1997). (A)

\item
``A Well-Collimated Quasi-Continuous Atom Laser,'' E.~W. Hagley,
L.~Deng, M.~Kozuma, J.~Wen, K.~Helmerson, S.~L. Rolston, and W.~D.
Phillips, Science {\bf 283}, 1706--1709 (1999). (I)

\item
``Radio-Frequency Output Coupling of the Bose-Einstein Condensate
for Atom Lasers,'' Y.~B. Band, P.~S. Julienne, and M.~Trippenbach,
Phys. Rev. A {\bf 59}, 3823--3831 (1999). (A)

\item
``Atom Laser with a CW Output Coupler,'' I.~Bloch, T.~W.
H{\"a}nsch, and T.~Esslinger, Phys. Rev. Lett. {\bf 82},
3008--3011 (1999). (I)

\item
``Optics with an Atom Laser Beam,'' I.~Bloch, M.~K{\"o}hl,
M.~Greiner, T.~W. H{\"a}nsch, and T.~Esslinger, Phys. Rev. Lett.
{\bf 87}, 030401/1--4 (2001). (I)

\item
``Measuring the Temporal Coherence of an Atom Laser Beam,''
M.~K{\"o}hl, T.~W. H{\"a}nsch, and T.~Esslinger, Phys. Rev. Lett.
{\bf 87}, 160404/1--4 (2001). (I)

\item
``Atom Laser Divergence,'' Y.~Le~Coq, J.~H. Thywissen, S.~A.
Rangwala, F.~Gerbier, S.~Richard, G.~Delannoy, P.~Bouyer, and
A.~Aspect, Phys. Rev. Lett. {\bf 87}, 170403/1--4 (2001). (I)

\eref

\subsection{Multiple Condensates}

\subsubsection{Different Internal States}

The alkali atoms (and hydrogen) have two hyperfine levels in their
ground states and numerous Zeeman sublevels (``spin states'').
Atoms in different internal states are distinguishable from one
another, and therefore can form independent (but interacting)
condensates.

Experimentally, there are several difficulties that arise. Atoms
in different hyperfine states tend to undergo exoergic
spin-exchange collisions that cause a great deal of heating and
atom loss. A fortuitous coincidence in singlet and triplet
scattering lengths suppresses these losses in ${}^{87}$Rb, making
long-lived double condensates possible.

Within the same hyperfine level, different Zeeman sublevels have
different magnetic moments, and magnetic traps tend to confine
them differently (or not at all). Optical traps, which confine
atoms by the interaction of an electric-field gradient and induced
electric-dipole moment, are one solution to this difficulty that
is finding increasing use (see Ref.[\ref{opticaltrap}]).

The physics of condensates in multiple internal states is quite
rich. Creation of a double condensate system from a single
condensate typically is done by introducing a resonant
electromagnetic field. Where interconversion of the spins is
possible, metastable spin domains can form. Topological phases can
be introduced by driving spatially-selective transitions from one
state to the other --- as was done in creating the first vortices
identified in a dilute-gas condensate (see
Ref.~[\ref{Matthews1999}]).

\bref

\item
``Local Spin-Gauge Symmetry of the Bose-Einstein Condensates in
Atomic Gases,'' T.-L. Ho and V.~B. Shenoy, Phys. Rev. Lett. {\bf
77}, 2595--2599 (1996). (I)

\item
``Binary Mixtures of {B}ose Condensates of Alkali Atoms,'' T.-L.
Ho and V.~B. Shenoy, Phys. Rev. Lett. {\bf 77}, 3276--3270 (1996).
(I)

\item
``Production of Two Overlapping {B}ose-{E}instein Condensates by
Sympathetic Cooling,'' C.~J. Myatt, E.~A. Burt, R.~W. Ghrist,
E.~A. Cornell, and C.~E. Wieman, Phys. Rev. Lett. {\bf 78},
586--589 (1997). (I)

\item
``Collisional Stability of Double Bose Condensates,'' P.~S.
Julienne, F.~H. Mies, E.~Tiesinga, and C.~J. Williams, Phys. Rev.
Lett. {\bf 78}, 1880--1883 (1997). (I)

\item
``\,`Stability Signature' in Two-Species Dilute Bose-Einstein
Condensates,'' C.~K. Law, H.~Pu, N.~P. Bigelow, and J.~H. Eberly,
Phys. Rev. Lett. {\bf 79}, 3105--3108 (1998). (I)

\item
``Dynamical Response of a {B}ose-{E}instein Condensate to a
Discontinuous Change in Internal State,'' M.~R. Matthews, D.~S.
Hall, D.~S. Jin, J.~R. Ensher, C.~E. Wieman, E.~A. Cornell,
F.~Dalfovo, C.~Minniti, and S.~Stringari, Phys. Rev. Lett. {\bf
81}, 243--247 (1998). (I)

\item
``Dynamics of Component Separation in a Binary Mixture of
{B}ose-{E}instein Condensates,'' D.~S. Hall, M.~R. Matthews, J.~R.
Ensher, C.~E. Wieman, and E.~A. Cornell, Phys. Rev. Lett. {\bf
81}, 1539--1542 (1998). (I)

\item
``Spin Domains in Ground-State {B}ose-{E}instein Condensates,''
J.~Stenger, S.~Inouye, D.~M. Stamper-Kurn, H.-J. Miesner, A.~P.
Chikkatur, and W.~Ketterle, Nature {\bf 396}, 345--348 (1998). (I)


\item
``Dynamics of Two Interacting Bose-Einstein Condensates,''
A.~Sinatra, P.~O. Fedichev, Y.~Castin, J.~Dalibard, and G.~V.
Shylapnikov, Phys. Rev. Lett. {\bf 82}, 251--254 (1999). (I)

\item
``Observation of Metastable States in Spinor {B}ose-{E}instein
Condensates,'' H.-J. Miesner, D.~M. Stamper-Kurn, J.~Stenger,
S.~Inouye, A.~P. Chikkatur, and W.~Ketterle, Phys. Rev. Lett. {\bf
82}, 2228--2231 (1999). (I)

\item
``Quantum Tunnelling across Spin Domains in a {B}ose-{E}instein
Condensate,'' D.~M. Stamper-Kurn, H.-J. Miesner, A.~P. Chikkatur,
S.~Inouye, J.~Stenger, and W.~Ketterle, Phys. Rev. Lett. {\bf 83},
661--665 (1999). (I)

\item
``Watching a Superfluid Untwist Itself: {R}ecurrence of {R}abi
Oscillations in a {B}ose-{E}instein Condensate,'' M.~R. Matthews,
B.~P. Anderson, P.~C. Haljan, D.~S. Hall, M.~J. Holland, J.~E.
Williams, C.~E. Wieman, and E.~A. Cornell, Phys. Rev. Lett. {\bf
83}, 3358--3361 (1999). (I)

\item \label{williams}
``Preparing Topological States of a {B}ose-{E}instein
Condensate,'' J.~E. Williams and M.~J. Holland, Nature {\bf 401},
568--572 (1999). (I)

\item
``Collective Oscillations of Two Colliding {B}ose-{E}instein
Condensates,'' P.~Maddaloni, M.~Modugno, C.~Fort, F.~Minardi, and
M.~Inguscio, Phys. Rev. Lett. {\bf 85}, 2413--2417 (2000). (I)

\item
``Skyrmions in a Ferromagnetic Bose-Einstein Condensate,''
U.~Al~Khawaja and H. Stoof, Nature {\bf 411}, 918--920 (2001). (I)

\eref

\subsubsection{Different Atoms}

Double condensates have been produced in a system of ${}^{41}$K
and ${}^{87}$Rb. Sympathetic cooling of the Rb atoms was used to
condense the K atoms; see also Ref.~[\ref{Modugno2001}].

\bref

\item
``Prospects for Mixed-Isotope Bose-Einstein Condensates in
Rubidium,'' J.~P. Burke,~Jr., J.~L. Bohn, B.~D. Esry, and C.~H.
Greene, Phys. Rev. Lett. {\bf 80}, 2097--2100 (1998). (I)

\item
``Two Atomic Species Superfluid,'' G.~Modugno, M.~Modugno,
F.~Riboli, G.~Roati, and M.~Inguscio, Phys. Rev. Lett. {\bf 89},
190404/1--4 (2002). (I)

\eref

\subsubsection{Atoms and Molecules}

A molecule is chemically distinct from the two atoms that it
comprises, and a process that produces molecules coherently in
fact should create a Bose-Einstein condensate of molecules. Two
approaches have been developed. The first is to create molecules
by photoassociation, essentially using a stimulated Raman
transition to place a pair of atoms into a particular molecular
state. The other method uses Feshbach resonances to produce
quasibound molecular states.

\bref

\item
``Superchemistry: Dynamics of Coupled Atomic and Molecular
Bose-Einstein Condensates,'' D.~J. Heinzen, R.~Wynar, P.~D.
Drummond, and K.~V. Kheruntsyan, Phys. Rev. Lett. {\bf 84},
5029--5033 (2000). (I)

\item
``Molecules in a {B}ose-{E}instein Condensate,'' R.~Wynar, R.~S.
Freeland, D.~J. Han, C.~Ryu, and D.~J. Heinzen, Science {\bf 287},
1016--1019 (2000). (I)

\item
``Formation of Pairing Fields in Resonantly Coupled Atomic and
Molecular Bose-Einstein Condensates,'' M.~Holland, J.~Park, and
R.~Walser, Phys. Rev. Lett. {\bf 86}, 1915--1918 (2001). (A)

\item
``Photoassociation of Sodium in a {B}ose-{E}instein Condensate,''
C.~McKenzie, J.~H. Denschlag, H.~H{\"a}ffner, A.~Browaeys,
L.~E.~E. de~Araujo, F.~K. Fatemi, K.~M. Jones, J.~E. Simsarian,
D.~Cho, A.~Simoni, E.~Tiesinga, P.~S. Julienne, K.~Helmerson,
P.~D. Lett, S.~L. Rolston, and W.~D. Phillips, Phys. Rev. Lett.
{\bf 88}, 120403/1--4 (2002). (I)

\item
``Atom-Molecule Coherence in a {B}ose-{E}instein Condensate,''
E.~A. Donley, N.~R. Claussen, S.~T. Thompson, and C.~E. Wieman,
Nature {\bf 417}, 529--533 (2002). (I)

\item
``Ramsey Fringes in a Bose-Einstein Condensate between Atoms and
Molecules,'' S.~J.~J.~M.~F. Kokkelmans and M.~J. Holland, Phys.
Rev. lett. {\bf 89}, 180401/1--4 (2002). (A)

\eref

\subsubsection{Bose-Fermi Mixtures}

Fermions obey the Pauli exclusion principle and are therefore
unable to form Bose-Einstein condensates without some sort of
pairing. Recent experiments have managed to achieve simultaneous
degeneracy in both Fermi and Bose gases, promising a whole new
realm of experiments and theory.

\bref

\item
``Observation of Fermi Pressure in a Gas of Trapped Atoms,'' A.~G.
Truscott, K.~E. Strecker, W.~I. McAlexander, G.~B. Partridge, and
R.~G. Hulet, Science {\bf 291}, 2570--2572 (2001). (I)

\item
``Quasipure {B}ose-{E}instein Condensate
 Immersed in a {F}ermi Sea,'' F.~Schreck, L.~Khaykovich, K.~L. Corwin, G.~Ferrari,
T.~Bourdel, J.~Cubizolles, and C.~Salomon, Phys. Rev. Lett. {\bf
87}, 080403/1--4 (2001). (I)

\item
``Two-Species Mixture of Quantum Degenerate {B}ose and {F}ermi
Gases,'' Z.~Hadzibabic, C.~A. Stan, K.~Dieckmann, S.~Gupta, M.~W.
Zwierlein, A.~G{\"o}rlitz, and W.~Ketterle, Phys. Rev. Lett. {\bf
88}, 160401/1--4 (2002). (I)

\item
``Fermi-Bose Quantum Degenerate ${}^{40}$K--${}^{87}$Rb Mixture
with Attractive Interaction,'' G.~Roati, F.~Riboli, G.~Modugno,
and M.~Inguscio, Phys. Rev. Lett. {\bf 89}, 150403/1--4 (2002).
(I)

\eref

\subsection{Lower Dimensions}

Bose-Einstein condensation is impossible in a uniform
one-dimensional gas, or even in a two-dimensional gas (except at
$T=0$). In a sufficiently tight potential, however, the healing
length $\xi=1/\sqrt{4\pi n a}$ may be larger than the condensate
extent in two or one dimensions and a quasi-1D or quasi-2D
condensate can be formed. In 1D, if the healing length becomes
smaller than the interparticle separation ($n^{-1/3}$), the Bose
gas becomes impenetrable; this is the Tonks gas, in which the Bose
gas begins to adopt fermionic properties.

\bref

\item
``The Complete Equation of State of One, Two and Three-Dimensional
Gases of Hard Elastic Spheres,'' L.~Tonks, Phys. Rev. {\bf 50},
955--963 (1936). Foundation paper. (I)

\item
``Existence of Long-Range Order in One and Two Dimensions,'' P.~C.
Hohenberg, Phys. Rev. {\bf  158}, 383--386 (1967). (I)

\item
``Bose-Einstein Condensation in Low-Dimensional Traps,''
V.~Bagnato and D.~Kleppner, Phys. Rev. A {\bf 44}, 7439--7441
(1991). (I)

\item
``Atomic Scattering in the Presence of an External Confinement and
a Gas of Impenetrable Bosons,'' M.~Olshanii, Phys. Rev. Lett. {\bf
81}, 938--941 (1998). (A)


\item
``Low-Dimensional Bose Liquids: Beyond the Gross-Pitaevskii
Approximation,'' E.~B. Kolomeisky, T.~J. Newman, J.~P. Straley,
and X.~Qi, Phys. Rev. Lett. {\bf 85}, 1146--1149 (2000). (A)

\item
``Bose-Einstein Condensation in Quasi-2D Trapped Gases,'' D.~S.
Petrov, M.~Holzmann, and G.~V. Shlyapnikov, Phys. Rev. Lett. {\bf
84}, 2551--2555 (2000). (A)

\item
``Regimes of Quantum Degeneracy in Trapped 1D Gases,'' D.~S.
Petrov, G.~V. Shlyapnikov, and J.~T.~M. Walraven, Phys. Rev. Lett.
{\bf 85}, 3745--3749 (2000). (A)

\item
``Realization of {B}ose-{E}instein Condensates in Lower
Dimensions,'' A.~G{\"o}rlitz, J.~M. Vogels, A.~E. Leanhardt,
C.~Raman, T.~L. Gustavson, J.~R. Abo-Shaeer, A.~P. Chikkatur,
S.~Gupta, S.~Inouye, T.~Rosenband, and W.~Ketterle, Phys. Rev.
Lett. {\bf 87}, 130402/1--4 (2001). (I)

\item
``Solutions of the Gross-Pitaevskii Equation in Two Dimensions,''
M.~D. Lee and S.~A. Morgan, J. Phys. B: At. Mol. Opt. Phys. {\bf
35}, 3009--3017 (2002). (I)

\eref

\subsection{Spectroscopy and Precision Measurement}

Ultracold atomic systems now routinely are interrogated
spectroscopically, since motional broadening of the transitions
can be extremely small. One might suspect that Bose condensates
would represent the ultimate limit in motional line narrowing. The
relatively high densities in condensates introduce mean-field
shifts to atomic resonances, however; these can depend
dramatically on the atomic density. The shifts have been used
effectively to help determine the onset of degeneracy in hydrogen
but present a problem for other precision measurements in which
the transition frequency is of primary interest.

\bref

\item
``Heisenberg-Limited Spectroscopy with Degenerate Bose-Einstein
Gases,'' P.~Bouyer and M.~A. Kasevich, Phys. Rev. A {\bf 56},
R1083--R1086 (1997). (I)

\item
``Cold Collision Frequency Shift of the $1S$--$2S$ Transition in
Hydrogen,'' T.~C. Killian, D.~G. Fried, L.~Willmann, D.~Landhuis,
S.~C. Moss, T.~J. Greytak, and D.~Kleppner, Phys. Rev. Lett. {\bf
81}, 3807--3811 (1998). (I)

\item
``Bragg Spectroscopy of a {B}ose-{E}instein Condensate,''
J.~Stenger, S.~Inouye, A.~P. Chikkatur, D.~M. Stamper-Kurn, D.~E.
Pritchard, and W.~Ketterle, Phys. Rev. Lett. {\bf 82}, 4569--4573
(1999); {\it ibid.} {\bf 84}, 2283 (2000). (I)

\item
``Contrast Interferometry using {B}ose-{E}instein Condensates to
Measure $h/m$ and $\alpha$,'' S.~Gupta, K.~Dieckmann,
Z.~Hadzibabic, and D.~E. Pritchard, Phys. Rev. Lett. {\bf 89},
140401/1--4 (2002). (I)

\eref

\subsection{Entanglement}

A quantum state of two or more particles that cannot be decomposed
into a direct product of single-particle states is said to be
entangled. Entanglement is therefore manifested in
quantum-mechanical correlations between the particles. When these
correlations exist in the momentum or position states of the
particles they can be responsible for the ``spooky action at a
distance'' that so bothered Einstein. Applications of entangled
states to quantum information processing are common today.

Entanglement begins with particles in a pure state. The
Bose-Einstein condensate, with its particles all in the
single-particle ground state, is therefore an auspicious starting
point. In these examples, binary collisions are used to create the
entanglement between the internal (spin) states and the external
(typically momentum) states.

\bref

\item
``Entanglement of Atoms via Cold Controlled Collisions,''
D.~Jaksch, H.~J. Briegel, J.~I. Cirac, C.~W. Gardiner, and
P.~Zoller, Phys. Rev. Lett. {\bf 82}, 1975--1978 (1999). (I)

\item
``Creating Macroscopic Atomic Einstein-Podolsky-Rosen States from
Bose-Einstein Condensates,'' H.~Pu and P.~Meystre, Phys. Rev.
Lett. {\bf 85}, 3987--3990 (2000). (A)

\item
``Squeezing and Entanglement of Atomic Beams,'' L.-M. Duan,
A.~S{\o}rensen, J.~I. Cirac, and P.~Zoller, Phys. Rev. Lett. {\bf
85}, 3991--3994 (2000). (A)

\item
``Many-Particle Entanglement with Bose-Einstein Condensates,''
A.~S{\o}rensen, L.-M. Duan, J.~I. Cirac, and P.~Zoller, Nature
{\bf 409}, 63--65 (2001). (A)

\item
``Creating Massive Entanglement of Bose-Einstein Condensed
Atoms,'' K.~Helmerson and L.~You, Phys. Rev. Lett. {\bf 87},
170402/1--4 (2001). (I)

\item
``Generation of Macroscopic Pair-Correlated Atomic Beams by
Four-Wave Mixing in {B}ose-{E}instein Condensates,'' J.~M. Vogels,
K.~Xu, and W.~Ketterle, Phys. Rev. Lett. {\bf 89}, 020401/1--4
(2002). (I)

\eref

\subsection{Cosmology}

It is curious that some of the most energetic processes in the
universe can be simulated with some of its least energetic
entities. Theoretical proposals suggest that acoustic or optical
analogues of black holes can be created using a condensate with a
vortex rotating faster than the local speed of sound or light,
respectively. Analogues of Hawking radiation also have been
proposed. Still another example is the collapse of condensates
with attractive interactions, which closely resembles the collapse
of stars in supernovae; see Ref.[\ref{Donley2001}].

\bref

\item
``Tabletop Astrophysics,'' P.~Ball, Nature {\bf 411}, 628--630
(2001). Review article. (E)

\item
``Relativistic Effects of Light in Moving Media with Extremely Low
Group Velocity,'' U.~Leonhardt and P.~Piwnicki, Phys. Rev. Lett.
{\bf 84}, 822--825 (2000). (A)

\item
``Bose-Einstein Condensates with $1/r$ Interatomic Attraction:
Electromagnetically Induced `Gravity' '', D.~O'Dell,
S.~Giovanazzi, G.~Kurizki, and V.~M. Akulin, Phys. Rev. Lett. {\bf
84}, 5687--5790 (2000). (A)

\item
``Sonic Analog of Gravitational Black Holes in Bose-Einstein
Condensates,'' L.~J. Garay, J.~R. Anglin, J.~I. Cirac, and
P.~Zoller, Phys. Rev. Lett. {\bf 85}, 4643--4647 (2000). (A)

\end{Reference}
\normalsize

\section{Acknowledgments}

The author would like to thank C.~W. Clark, R.~G. Hulet,
K.~Jagannathan, W.~Ketterle, E.~A. Newman, and R.~H. Romer for
helpful comments and suggestions. This work was supported in part
by NSF grant no. PHY-0140207.



\end{document}